\documentclass[prd,11pt,onecolumn,tightenlines,nofootinbib,groupedaddress]{revtex4}
\usepackage{amsmath,amsfonts}
\usepackage{subfigure}
\usepackage{enumerate}
\usepackage{braket}
\usepackage[colorlinks,citecolor=blue,linkcolor=blue,urlcolor=blue]{hyperref}
\usepackage{graphicx}
\usepackage{titletoc,titlesec}
\usepackage[T1]{fontenc}

\newcommand{\M}{{\mathbb M}}
\newcommand{\R}{{\mathbb R}}

\newcommand{\va}{\scriptscriptstyle}

\newcommand{\D}{\Delta}
\newcommand{\be}{\begin{equation}}
\newcommand{\ee}{\end{equation}}
\DeclareMathAlphabet{\mathfs}{U}{rsfs}{m}{n} 
\newcommand{\mfs}[1]{\mathfs {#1}}         
\newcommand{\s}{{\kappa_{\va SG}}}
\newcommand{\rH}{r_{\va H}}
\newcommand{\rO}{r_{\va O}}
\newcommand{\Mi}{{\va M}}

\newcommand{\xBH}{x_{\va BH}}
\newcommand{\xo}{x_{\va 0}}
\newcommand{\scri}{{\mfs J}}
\newcommand{\sL}{{\mfs L}}
\newcommand{\sI}{{\mfs I}}

\def\D{\Delta}

\newcommand{\sg}{{\kappa_{\va SG}}}

\renewcommand\thesection{\Roman{section}}
\titleformat{\section}{\large\scshape\bfseries\centering}{\thesection.}{.7em}{}
\titleformat{\subsection}{\scshape\bfseries}{\thesubsection.}{.7em}{}
\titleformat{\subsubsection}{\scshape}{\thesubsubsection.}{.7em}{}

\begin{document}

\title{\Large\scshape\bfseries Light Cone Black Holes}

\author{Tommaso De Lorenzo$^{1,2}$}
\email{tdelorenzo@psu.edu}
\author{Alejandro Perez$^1$}
\email{perez@cpt.univ-mrs.fr}

\affiliation{$^1$Aix Marseille Univ, Universit\'e de Toulon, CNRS, CPT, 13000 Marseille, France \vspace{1ex}\\$^2$Institute for Gravitation and the Cosmos \& Physics Department, Penn State, University Park, PA 16802, U.S.A.}

\begin{abstract}
When probed with conformally invariant matter fields, light cones in Minkowski spacetime satisfy thermodynamical relations which are the analog of those satisfied by stationary black holes coupled to standard matter fields. These properties stem from the fact that light cones are conformal Killing horizons stationary with respect to observers following the radial conformal Killing fields in flat spacetime. The four laws of light cone thermodynamics relate notions such as (conformal) temperature, (conformal) surface gravity,  (conformal) energy and a conformally invariant notion related to area change. These quantities do not admit a direct physical interpretation in flat spacetime. However, they become the usual thermodynamical quantities when Minkowski is mapped, via a Weyl transformation, to a target spacetime where the conformal Killing field becomes a proper Killing field. In this paper we study the properties of such spacetimes. The simplest realisation turns out to be the Bertotti-Robinson solution, which is known to encode the near horizon geometry of near extremal and extremal charged black holes. The analogy between light cones in flat space and black hole horizons is therefore strengthened. The construction works in arbitrary dimensions; in two dimensions one recovers the Jackiv-Teitelboim black hole of dilaton gravity.
Other interesting realisations are also presented.
\end{abstract}

\maketitle
\tableofcontents
\section{Introduction}

\begin{figure}[!h]
\begin{center}
\hspace{.6cm}
\begin{minipage}[c]{.33\textwidth}
\centering
	\includegraphics[height=8cm]{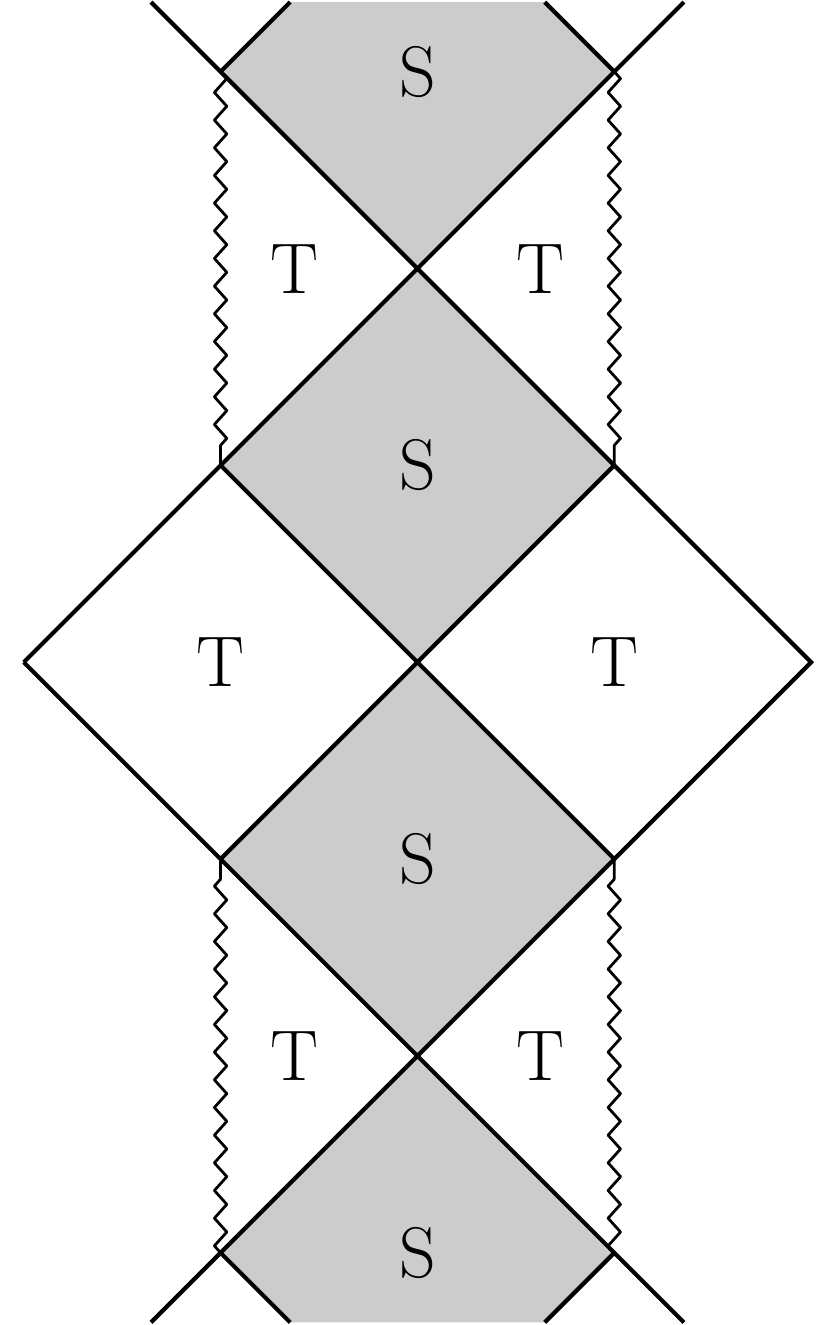}
\end{minipage}%
\begin{minipage}[c]{.6\textwidth}
\centering
          \hspace{.6cm}\includegraphics[height=8cm]{light-cones}
\end{minipage}\\
\vspace{2cm}
\hspace{.6cm}
\begin{minipage}[c]{.33\textwidth}
	\hspace{.6cm}
	\includegraphics[height=8cm]{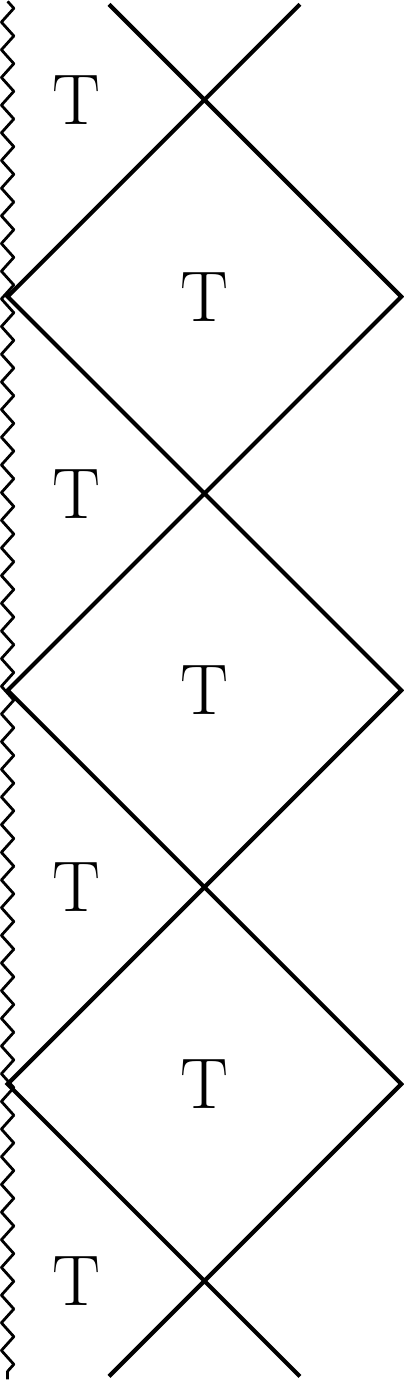}
\end{minipage}
\begin{minipage}[c]{.6\textwidth}
\centering
         \hspace{.6cm} \includegraphics[height=8cm]{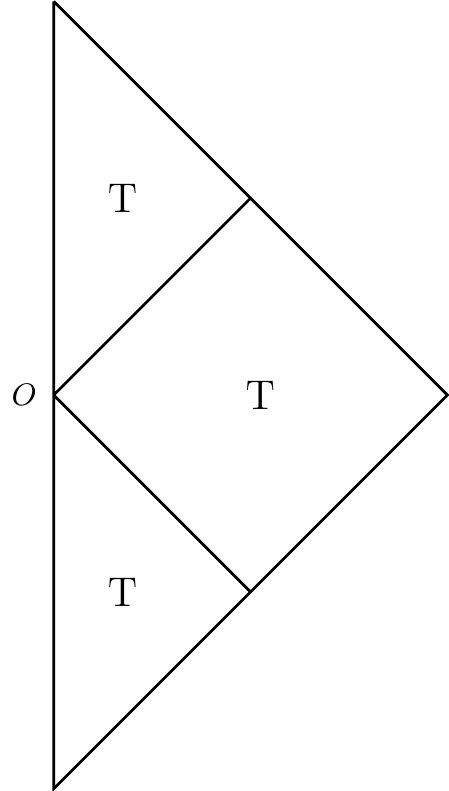}
\end{minipage}
\end{center}

\caption{The Penrose diagram of the Reissner-Nordstrom black hole on the left compared with the causal structure of the radial CKF in Minkowski spacetime on the right, in both the non-extremal $\Delta>0$ and extremal $\Delta =0$ case. The letters $S$ and $T$ designate the regions where the Killing or conformal Killing fields are spacelike or timelike respectively. The light cone emanating from the points $O^\pm$ (and $O$ in the extremal case) are the hypersurface where the MCKF is null.
This figure is taken from \cite{DeLorenzo:2017tgx}.}
\label{fig:penrose}
\end{figure}

The in-going and out-going light surfaces emanating from a sphere of radius $\rH$ at time $t$ in Minkowski spacetime is a bifurcate conformal Killing horizon \cite{RCKF}. The associated conformal Killing vector field becomes null on the light cones of the two events where the previous null surfaces converge. These null surfaces separate the whole of Minkowski spacetime in regions where the conformal Killing field $\xi^a$ is either timelike or spacelike. These regions are in formal correspondence with the different regions defined by the outer and inner horizons of non-extremal Reissner-Nordstrom black holes. When $\rH\to 0$ some regions collapse and the causal features of the conformal Killing field now correspond to those of extremal Reissner-Nordstrom black holes (see Figure \ref{fig:penrose}).

 This resemblance is actually more profound than what it might seem at first sight. It was shown in \cite{DeLorenzo:2017tgx}, that the previous light cone surfaces satisfy thermodynamical properties that are analogous to those of black holes when tested or perturbed with conformally invariant matter. Here we will show that Minkowski spacetime can be mapped via a Weyl transformation to target spacetimes where the conformal Killing field becomes a proper Killing field, and the associated light cones turn into Killing horizons. There is a certain freedom in the choice of the conformal map which leads to different geometric features of the Killing horizon in the target spacetime. In the most natural case we will show that the target spacetime represents the near horizon near extremal approximation of Reissner-Nordstrom black holes, shedding light on the profound resemblance between Minkowski light cones and black holes. Other natural realisations of such target spacetimes are studied. Because the thermodynamical notions entering the analysis of  \cite{DeLorenzo:2017tgx} are all conformally invariant quantities, the mapping of the light cones to Killing horizons via a Weyl transformation clarifies, in this way, their intrinsic physical meaning.

In order to introduce the present work let us first briefly  and partially review the analysis of  \cite{DeLorenzo:2017tgx}. Minkowski Conformal Killing Fields (MCKFs) define conformal bifurcating Killing horizons. They carry a conformally invariant  \cite{Jacobson:1993pf} notion of surface gravity $\kappa_{\va SG}$ defined by the following equation
\be\label{susu}
\nabla_a (\xi\cdot \xi)\,\hat=-2\sg  \xi_a \, , 
\ee
where $\hat=$ denotes an equality that holds at the conformal Killing horizon. All four laws of black hole thermodynamics have a suitable version for conformal Killing horizons defined by MCKFs:  The surface gravity $\kappa_{\va SG}$ is constant on the horizon and it is associated to a mathematical notion of temperature (conformal temperature)  \be \label{zero}{T=\frac{\kappa_{\va SG}}{2\pi}={\rm constant}}.\ee 
 The relation between $\kappa_{\va SG}$ and conformal temperature can be established in different ways. In \cite{DeLorenzo:2017tgx} this was done via the standard Bogoliubov transformation relating inertial and suitable asymptotic conformal observers in flat spacetime, and also recovered via a Wick rotation to Euclidean time 
(all standard in quantum field theory). We will revisit the second procedure in a very general fashion in Section \ref{CT}.

Under linear perturbations induced by conformally invariant matter fields the current $J_a=\delta T_{ab}\xi^b$ is conserved, namely
\be\label{cons}
\nabla^a J_a=\nabla^a (\delta T_{ab} \xi^b)=0.
\ee
The previous equation can be used to establish a suitable version of the first law for MCKF 
\be\label{firstLaw}
{\delta M = \frac{\kappa_{\va SG}}{8\pi} \, {\delta A}+\delta M_{\infty}\,,}
\ee
where
\be\label{M}
\delta M = \int_\Sigma J_a d\Sigma^a
\ee
is the conformally invariant mass of the perturbation evaluated at an initial Cauchy surface $\Sigma$, $\delta M_{\infty}$ is the conformal mass flow at $\sI^+$, and $\delta A$ is a conformally invariant notion defined as
\be \label{eq:area}
 \delta A \equiv \int\limits_{H^+} \frac{\kappa}{\s} \,\delta \theta dS^2 dv.
\ee
Here $\delta \theta$ is the first order perturbation of the expansion of the generator of the horizon, $v$ is the advanced Minkowski time (a natural affine parameter for the generators), $dS^2$ the flat background area measure of the spherical cross section $v=$constant of the unperturbed light cone, and $\kappa$ is defined by
\be\label{eq:inaffinity}
 \xi^a \nabla_a \xi^b {\,\hat=\,}\kappa\,\xi^b\,.
\ee
Unlike $\kappa_{\va SG}$, the function $\kappa$ is not constant and is not conformally  invariant. 
For a proof of the conformal invariance of the quantities involved in the first law (\ref{firstLaw}) see   \cite{DeLorenzo:2017tgx}.

Provided that $\delta T_{ab}$ satisfies the weak energy condition (equivalent to the strong energy condition for conformally invariant matter), the second law holds, namely
\be\label{area}
{\delta A\ge 0.}
\ee
 
 Finally, in the ``extremal'' limit $\rH\to 0$ the temperature (\ref{zero}) goes to zero as well as the area of the bifurcate sphere $A=4\pi \rH^2$. We can interpret this as a form of third law of thermodynamics. 
This, plus equations (\ref{zero}), (\ref{firstLaw}), and (\ref{area}), are the analog of the laws of black hole mechanics for the light cones in flat spacetimes that define the conformal Killing horizons associated to MCKFs \cite{DeLorenzo:2017tgx}. 

The previous formal analogy between the properties of MCKFs and thermodynamics of black holes is interesting because it captures some basic mathematical features on a  background with trivial gravitational field. However, the reserve side of it is that the various conformal invariant notions entering the laws have no clear physical meaning: $\delta M$ is not an energy measured by any real physical observer, the conformal temperature is not the one detected by any physical thermometer (for more discussion see  \cite{DeLorenzo:2017tgx}), and $\delta A$ is related to area change (due to the presence of the perturbation of the expansion $\delta \theta$ in equation (\ref{eq:area})) but, as it stands, does not correspond to any direct geometric notion of area change of the bifurcating sphere. This is so due to the fact that $\kappa_{\va SG}/\kappa\not=1$ for conformal Killing fields in general and  the MCKFs in particular.

Nevertheless, the previous thermodynamical analogy can be more clearly understood if one performs a conformal transformation sending $(\R^4, \eta_{ab})$ to a target spacetime $(M,g_{ab})$ with $ g_{ab}=\omega^2 \eta_{ab}$ so that $\xi^a$ becomes a genuine Killing field. The conformal bifurcate horizons turn consequently into bifurcate Killing horizons in the target spacetime. If $\delta T_{ab}$ comes from a conformally invariant matter model then $\delta \tilde T_{ab}=\omega^{-2} \delta T_{ab}$ and the conservation of the associated current $\tilde J_a$, equation \eqref{cons} with tilded quantities, holds  in the new spacetime \cite{wald2010general}. Equations (\ref{zero}), (\ref{firstLaw}), and (\ref{area}) remain true in the target spacetime with identical numerical values for a given perturbation. However, all the quantities involved acquire now the standard physical and geometric meaning that they have in the context of black holes.

The question we want to explore here is what are the generic global features of the spacetimes obtained by the previous procedure. Are there cases where these spacetimes represent black holes? What are they in the other cases? 

There is clearly an infinite number of possibilities. Indeed, if  $\omega_1$ defines a Weyl transformation with the previous desired properties, then $\omega_2$ defines a new suitable Weyl transformation as long as $\xi(\omega_2/\omega_1)=0$.  We will see that the generic global features can be made apparent in a small number of representative cases. The simplest case corresponds to $\omega\propto1/r^2$ (for $r$ a Minkowski radial coordinate) and it reproduces the Bertotti-Robinson solution \cite{Bertotti:1959pf, Robinson:1959ev} of Einstein-Maxwell theory---see Section~\ref{BBRR}. Such solution has been known to encode the near horizon geometry of close-to-extremal and extremal Reissner-Nordstrom black holes. 
Another representative example is the {de Sitter} realisation where the bifurcating horizons correspond to intersecting cosmological horizons (there is no black hole in this case)---see Section~\ref{sub:desitter}. Weakly asymptotically {(Anti)-de Sitter} black hole realisations are also presented---Section~\ref{sub:asyAdS}---, together with a more exotic asymptotically flat spacetime with Killing horizons but no black holes---Section~\ref{originalone}. 
 
Radial MCKF with conformal Killing horizons associated to light cones bifurcating at a sphere generalize to arbitrary dimensions. As long as the matter perturbing the geometry is conformally invariant,  the generalization of equations (\ref{zero}), (\ref{firstLaw}), and (\ref{area}) is also valid. The Bertotti-Robinson representation, which in arbitrary dimensions is given by $AdS_2\times S_{d-2}$, remains the simplest one.
For $d=2$ the light cone black hole corresponds to the Jackiw-Teitelboim solution \cite{jackiw1984quantum} of dilaton gravity.

\section{Radial Conformal Killing Fields in Minkowski Spacetime}
 
Consider Minkwoski spacetime in spherical coordinates
\be
\begin{split}
ds^2_\Mi = \eta_{\mu \nu}dx^\mu dx^\nu &= -dt^2 + dr^2 + r^2 d\Omega^2\\
&= -dvdu + \frac{(v-u)^2}{4} d\Omega^2
\end{split}
\ee
where $d\Omega^2$ is the unit-sphere metric, while $v=t+r$ and $u=t-r$ the standard Minkowskian null coordinates. The conformal group in four dimensional Minkowski spacetime $\M^4$  is isomorphic to the group $SO(5,1)$. Any generator defines a Conformal Killing Field in Minkowski spacetime (MCKF), namely a vector field $\xi$ along which the metric $\eta_{ab}$ changes only by a conformal factor:
\be\label{killing}
\sL_\xi \,\eta_{ab}=\nabla_a\xi_b+\nabla_b\xi_a=\frac{\psi}{2} \eta_{ab}
\ee
with
\be\label{calabaza}
\psi = \nabla_a \xi^a\,.
\ee
The 15 generators of $SO(5,1)$ are given in Euclidean coordinates $(t,x,y,z)$ by \cite{francesco2012conformal}
\be\label{siete}
\begin{split}
& P_\mu=\partial_{\mu} \ \ \ \ \ \ \ \ \ \ \ \ \ \ \ \ \ \ \ \ \ \ \ \ \ \ \ \ \ \ {\rm Translations} \\
& L_{\mu\nu}=\left(x_\nu\partial_{\mu} -x_\mu\partial_{\nu}\right)  \ \ \ \ \ \ \ \ \ \ \ \ \ {\rm Lorentz\ transformations} \\
& D= x^\mu\partial_{\mu}  \ \ \ \ \ \ \ \ \ \ \ \ \ \ \ \ \ \  \ \ \ \ \ \ \ \ \ \  {\rm Dilations}\\
& K_{\mu}=\left(2 x_\mu x^\nu \partial_{\nu} -x\cdot x\, \partial_{\mu}\right)\ \ \ \ \ \ \, {\rm Special\ conformal\ transformations,} 
\end{split}
\ee
where $f\cdot g \equiv f^\mu g_\mu$.
Dilations can be written as
\be
D=r\partial_r+t\partial_t
\ee
and $K_0$ as
\be
K_0=-2t D-(r^2-t^2) P_0\,.
\ee
Together with $P_0=\partial_t$, those are the only generators that do not contain angular components. Hence the most general purely radial MCKF has the form 
\be \label{explicit}
\begin{split}
\xi =&-a K_0+b D+c P_0\\
 =&(2at+b) D+ [a(r^2-t^2)+c]  P_0, 
\end{split}
\ee
with $a,b,c$ arbitrary constants. In terms of null Minkowski coordinates one has \cite{DeLorenzo:2017tgx}

\be\label{eq:CKF}
\xi^\mu \partial_\mu=(av^2+bv+c)\partial_v+(au^2+bu+c)\partial_u\,,
\ee
where $v=t+r$, $u=t-r$. Therefore, radial conformal Killing fields in flat spacetime are completely characterized by a single quadratic polynomial.

\begin{figure}[t]
\center
\includegraphics[height=9cm]{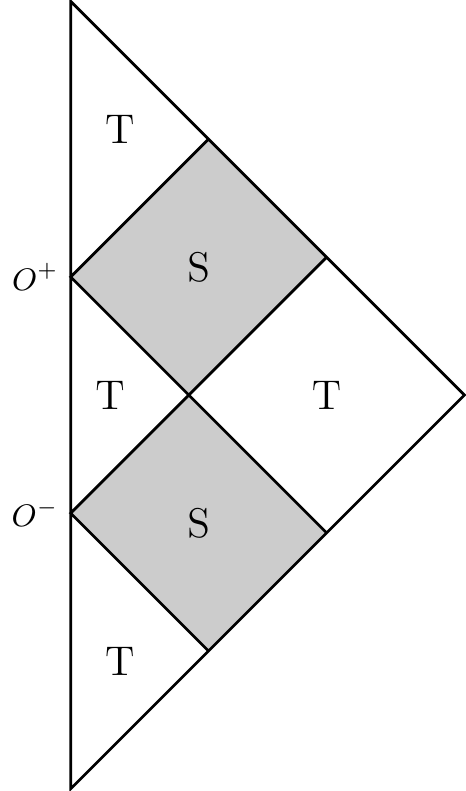} 
\caption{The most general radial conformal Killing field $\xi$ in Minkowski spacetime divides the latter in six regions. The field $\xi$ is spacelike in the shaded regions, timelike elsewhere. It becomes null on the light cones separating the regions. It vanishes at the tips of the ligthcones and at their intersection.}
\label{fig:Penrose_CD2}
\end{figure}

Radial MCKFs containing a bifurcating conformal Killing horizon are those conformal Killing fields for which $a\not=0$. Up to Poincar\`e transformations, they can be written as 
\be\label{killingg}
\xi^\mu \frac{\partial}{\partial x^\mu} 
=\frac{(v^2-r_{\va H}^2)}{r_{\va O}^2-r_{\va H}^2}\,\frac{\partial}{\partial v} + \frac{(u^2-r_{\va H}^2)}{r_{\va O}^2-r_{\va H}^2}\,\frac{\partial}{\partial u}
\ee
where $\rH$ and $\rO$ are two constants defined as follows. The first is the  radius $\rH$ of the  sphere at $t=0$ where the Killing field is zero $\xi|_{r=r_{\va H}}=0$. i.e., the bifurcation surface. The second quantity is the radius $\rO>\rH$ of a sphere on the surface $t=0$ where the Killing field is normalised, $\xi\cdot \xi|_{r=r_{\va O}}=-1$. For these choices, the norm of $\xi^a$ is given by
\be
\xi \cdot \xi = -\frac{(v^2-r_{\va H}^2)(u^2-r_{\va H}^2)}{(r_{\va O}^2-r_{\va H}^2)^2},
\ee
which reproduces the causal pattern illustrated in Figure \ref{fig:Penrose_CD2}. The regions where $\xi$ is time-like or space-like is denoted by $T$ and $S$ respectively with $\xi$ null-like at cones $u=\pm \rH$ and $v=\pm \rH$. The scalar $\psi$ defined in (\ref{calabaza}) is in this case given by
\be\label{bichito}
\psi=\frac{4(u+v)}{\rO^2-\rH^2}.
\ee

There is a subgroup $SL(2,\R)\subset SO(5,1)$ corresponding to the conformal group restricted to the $(r-t)$- ``plane''. This subgroup plays a role in the analysis of the relationship with near extremal black holes and their near horizon geometry that we will describe in more detail in Section \ref{nhs}.  The generators are
\be
\begin{split}
& T_0=\frac{1}{2} (r_{\va H} P_0-\frac{K_0}{\rH}) \\
& T_1=\frac{1}{2} (r_{\va H} P_0+\frac{K_0}{\rH})\\
& T_2=D,
\end{split}
\ee
satisfying the algebra 
\be\label{lieSL} 
[T_a,T_b]=\epsilon_{abc}\eta^{cd}T_d.
\ee In terms of these generators the vector field $\xi^a$ in \eqref{killingg} is simply
\be
\xi=-\frac{2\rH}{r_{\va O}^2-r_{\va H}^2} T_1.
\ee
One can arrange for linear combinations of $T_a$ to vanish either at $(v=\rH, u=\rH)$ (which includes the past outer horizon)  or at $(v=-\rH, u=-\rH)$ (which includes the future outer horizon). 

\subsection{The action of the conformal group and the Rindler wedge from MCKFs}

Rindler horizons are related to our radial MCKF by the action of an element of the conformal group generated via the exponentiation of (\ref{siete}). The  conformal group maps a radial MCKF into a MCKF that is not necessarily radial. To see this in a more precise way let us discuss the action of the various kinds of generators. The action of the Poincar\'e subgroup is very clear: it simply maps the events $O_{\pm}$ to new events in Minkowski spacetime in term of which the new spacetime diagram analogous to the one in Figure \ref{fig:Penrose_CD2} can be constructed from the light cones emanating from the new $O_{\pm}$. It is clear that the new events $O_{\pm}$ need not remain of the $r=0$ axis and can be transformed to arbitrary events on the flat spacetime. Pure dilations with parameter $\lambda$ send a radial MCKF to a new radial MCKF with rescaled parameters $(\rH, \rO)\to (\lambda \rH, \lambda \rO)$. As expected,  the radius of the bifurcate sphere and the observer sphere get rescaled. Finally, finite special conformal transformations are characterised by a four vector $b^{\mu}$. They are diffeomorphisms $x^{\mu}\to x^{\prime \mu}$, sending the Minkowski metric $\eta_{ab}\to \omega(x)^2 \eta_{ab}$ 
with 
\be
\omega(x)=(1-2b\cdot x+(b\cdot b) (x\cdot x)).
\ee
Their action can be expressed as
\be
\frac{x^{\mu}}{x\cdot x}=\frac{x^{\prime \mu}}{x^\prime \cdot x^\prime }-b^\mu.
\ee
Therefore, special conformal transformations can be viewed as the composition of an inversion, followed by a translation, and a second inversion \cite{francesco2012conformal}. Generically, they deform the radial MCKF into a non radial one.
An especially interesting situation arises when $b^0=0$: The first inversion sends the bifurcate sphere into a new sphere.  One can choose the translation $b^\mu$ so that a point on the sphere is shifted (in the intermediate translation) to the origin ($r=0$).  In this case, the second inversion sends that special point to spacial infinity and the rest of the sphere to a plane. The domain of dependence of the bifurcate sphere is mapped to the Rindler wedge \footnote{The causal complement (the outside of the diamond) is mapped to the complementary Rindler wedge but it does not cover it entirely, as simple topological considerations show (the outside is not simply connected in Minkowski spacetime). In order to get the entire complementary Rindler wedge one needs to add points that are beyond infinities $\sI_{\pm}$ in, for instance, the conformal compactification on the Einstein universe. There, the causal complement of the central diamond is itself a (now simply connected) diamond.} and the radial MCKF become the boost Killing vector field of Minkowski spacetime \cite{Martinetti:2002sz, Martinetti:2008ja}.  
 All this implies that the thermodynamical laws of Rindler horizons for conformal matter \cite{Bianchi:2013rya} are a special case of the laws derived in \cite{DeLorenzo:2017tgx}.

\subsection{From a radial MCKF to a Killing field on a target spacetime}

We just discussed how the radial MCKF that is timelike inside the ``diamond'' (domain of dependence of the sphere of radius $\rH$) can be mapped to the boost Killing field in Minkowski spacetime that defines the orbits of Rindler observers with their associated Rindler Killing horizon (the boundary of the Rindler wedge). These are the only Killing horizons in flat spacetime. In order to obtain more general Killing horizons with the causal features emphasised in Figure \ref{fig:Penrose_CD2},  one has to leave the realm of flat spacetimes by introducing more general  Weyl transformations preserving the causal structure but not necessarily the flatness condition present in the conformal group. Here we investigate  the possibility of turning the radial MCKF into a Killing field by mapping  Minkowski spacetime to a target curved geometry via general Weyl transformations.   

More precisely, consider any spacetime conformally related to Minkowski
\be
g_{ab} = \omega^2 \,\eta_{ab}\,.
\ee
In such a spacetime, the conformal Killing field $\xi$ remains so. Indeed
\be
\sL_{\xi}g_{ab}=\sL_{\xi}(\omega^2 \eta_{ab})=\left(\frac{\psi}{2} + \xi^a\partial_a \, (\log \,\omega^2)\right)g_{ab}.
\ee
In particular, there exist conformal transformations such that $\xi$ becomes a proper Killing field. From the above equation, it follows that those are given by conformal factors satisfying
\be\label{eq:conftokill}
\frac{\psi}{2} + \xi^a\partial_a \, (\log\omega^2)=0
\ee
Before solving this equation it is convenient to introduce  new  coordinates given by
\be\label{eq:CooTra}
\begin{split}
\tau&= \frac{(\rO^2-\rH^2)}{4\,\rH}\log\frac{(u-\rH)(v-\rH)}{(u+\rH)(v+\rH)}\\
x &=  \frac{2(\rH^2-u\, v)}{v-u}.
\end{split}
\ee
The Minkowski line element $ds_\Mi^2$ becomes
\be\label{eq:RN}
ds_\Mi^2=\frac{r^2}{\xo^2} \left(-\frac{x^2-x_{\va BH}^2}{\xo^2}\,d\tau^2+\frac{\xo^2}{x^2-x_{\va BH}^2} \, dx^2+x_{\va 0}^2 \,d\Omega^2\right)\,.
\ee
where
\be
\xBH\equiv {2 \,\rH},\ \ \ \xo^2\equiv\rO^2-\rH^2,
\ee
and where $r$ is now a function of $\tau$ and $x$ given by
\be\label{erre}
   r=\frac{\xBH^2}{2}\frac{\sqrt{x^2-
   \xBH^2} \cosh \left(\frac{\xBH}{\xo^2}\tau\right)+x}{
   \xBH^2\cosh ^2\left(\frac{\xBH}{\xo^2}\tau\right)-x^2 \sinh
   ^2\left(\frac{\xBH}{\xo^2}\tau\right)}
  \ee
In these coordinates, our radial MCKF reduces to 
\be\label{eq:killtau}
\xi^\mu \frac{\partial}{\partial x^\mu} =  \frac{\partial}{\partial \tau} \,,
\ee
and equation \eqref{eq:conftokill} becomes simply
\be\label{rescaling}
\frac{\partial}{\partial \tau} (\log \omega^2) = -\frac\psi2 \,.
\ee
Using the explicit value of $\psi$ given on \eqref{bichito}, the solution is
\be\label{question}
\log(\omega^2) = \log\left(\frac{\xo^2}{r^2}\right) + F(x,\theta,\varphi)\,,
\ee
where $F(x,\theta,\phi)$ is a general dimensionless function of $(x,\theta,\phi)$ and the Minkowski radial coordinate $r$ is the function of $\tau$ and $x$ given in \eqref{erre}.
Redefining for convenience $F \equiv -\log\,P^2$, we find 
\be\label{eq:omega}
\omega^2 = \frac{\xo^2}{r^2}\,\frac{1}{P^2(x,\theta,\varphi)}\,.
\ee
We could have directly guessed the answer from the form of the metric (\ref{eq:RN}) as $\xi^a$ is clearly a Killing field of the factor between brackets.
Notice that inspection of (\ref{eq:RN}) leads to the conclusion that the outer and inner light cone horizons are located respectively at $x_\pm=\pm \xBH$. Asymptotic null infinities $\scri^\pm$ are located at finite $x$, while space-like infinity $i^0$ is at $x \to +\infty$. Finally, the origin $u=v$ corresponds to $x \to {\rm sign}(\rH^2-t^2)\,\infty$.

The  function $P^2(x,\theta,\varphi)$  labels the members of an infinite family of Weyl transformations of Minkowski spacetime such that the target spacetime admits a genuine Killing field corresponding to radial MCKF \eqref{killingg}. The metric of such spacetimes is given by
\be\label{22}
ds^2=\omega^2 ds_\Mi^2=\frac{1}{P^2(x,\theta,\varphi)} \left(-\frac{x^2-x_{\va BH}^2}{\xo^2}\,d\tau^2+\frac{\xo^2}{x^2-x_{\va BH}^2} \, dx^2+x_{\va 0}^2 \,d\Omega^2\right)\,.
\ee
 Clearly, any additional coordinates transformation that does not depend on $\tau$ sends the metric in an equivalent $\tau$-independent form. In Appendix~\ref{CooTra} some interesting examples are presented. In particular, there we show the precise relation between the above coordinates and those used in  \cite{DeLorenzo:2017tgx}.

\subsection{The Hartle-Hawking temperature}\label{CT}

\begin{figure}[t]
\center
\includegraphics[height=8.3cm]{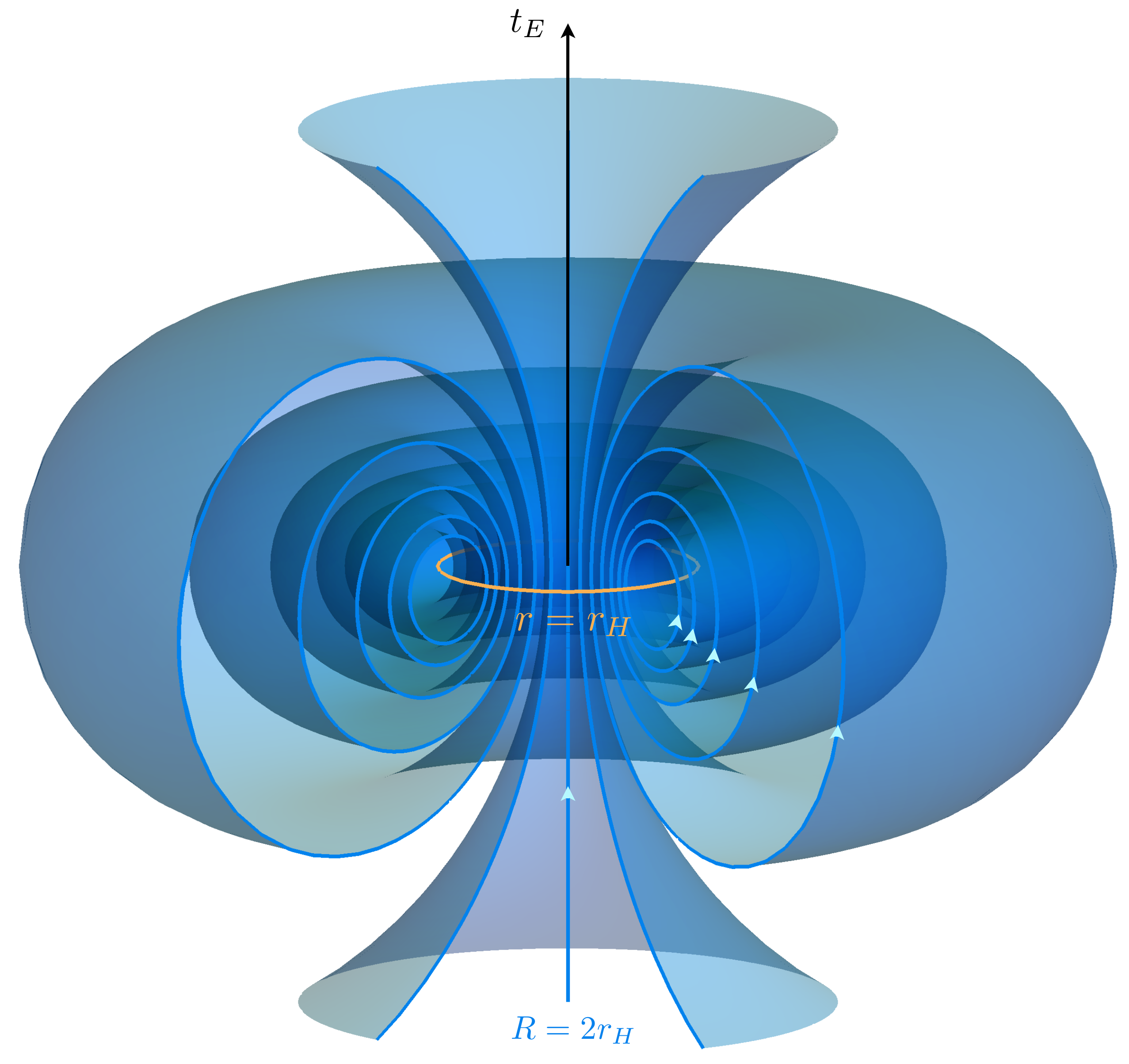}
\caption{Three dimensional representation of the flow of the conformal Killing field in the Euclidean spacetime $\R^4$. The orbits in this one-dimension-less representation are non-concentric tori around the bifurcate sphere $r=r_H$---here represented as a circle. They degenerate into the $t_E$ axis for $R=2r_H$. This figure is taken from \cite{DeLorenzo:2017tgx}.}
\label{fig:Euclidean}
\end{figure}

As we mentioned in the introduction, there is a conformally invariant notion of surface gravity \eqref{susu} and conformally invariant temperature \eqref{zero} associated to the light cone conformal Killing horizons that we are studying here. The simplest way to make such structure apparent is by introducing a Wick rotation in terms of the conformal Killing parameter $\tau$ defined in the previous section. 
Explicitly,  the standard Wick rotation $\tau \to -i \bar\tau$ turns the Minkowski metric into the Euclidean flat metric
\be\label{eq:RNE}
ds_E^2=\frac{r^2(-i\bar\tau,x)}{\xo^2} \left(\frac{x^2-x_{\va BH}^2}{\xo^2}\,d\bar\tau^2+\frac{\xo^2}{x^2-x_{\va BH}^2} \, dx^2+x_{\va 0}^2 \,d\Omega^2\right)\,.
\ee
Notice that under such transformation the Minkowski radial coordinate $r(-i\bar\tau,x)$ remains real---see \eqref{erre}---and therefore \eqref{eq:RNE} is a real euclidean metric  \cite{DeLorenzo:2017tgx}.
The same conclusion holds for any of the conformally related metrics \eqref{22} which under the Wick rotation become
\be\label{marche}
ds^2=\frac{1}{P^2(x,\theta,\varphi)}\left( \frac{x^2-x_{\va BH}^2}{\xo^2}\,d\bar\tau^2+\frac{\xo^2}{x^2-x_{\va BH}^2} \, dx^2+x_{\va 0}^2 \,d\Omega^2\right)\,.
\ee
The apparent singularity at $x=\xBH$ for the previous metrics can be removed in the usual way by introducing a new set of coordinates
\be
\begin{split}
\rho^2&=\frac{\xBH^2}{\xo^2}(x^2-\xBH^2)\\
\varphi&=\frac{\xBH}{\xo^2}\,\tau\,,
\end{split}
\ee
in terms of which the family of metrics \eqref{marche} becomes
\be\label{brk}
ds^2= \frac{1}{P^2}\left(\rho^2 d\varphi^2+\frac{\xo^2}{\xo^2+\rho^2}\ d\rho^2+\xo^2 d\Omega^2\right)\,.
\ee
Assuming that $P$ is non vanishing at $\rho=0$, the previous metrics  would have a conical singularity at $x=+\xBH$ unless $0\le\varphi\le 2\pi$.
Therefore, the quantum state of fields compatible with the topology of the Euclidean continuation must be a thermal state with temperature \be\label{pepino}
T_{\va HH}=\frac{1}{2\pi}\frac{\xBH}{\xo^2}=\frac{\rH}{\pi (\rO^2-\rH^2)}
\ee
which is the Hartle-Hawking temperature found in \cite{DeLorenzo:2017tgx}. In Figure \ref{fig:Euclidean} the Euclidean continuation of the radial MCKF is represented and shows its clear thermal features encoded in the closed nature of its orbits. This temperature---which we termed conformal temperature---is conformally invariant and it is related to the conformally invariant notion of surface gravity defined in the Introduction via the standard relation $T_{\va HH}=\kappa_{\va SG}/(2\pi)$.  In the  following sections  we will study the spacetime realisations corresponding to different choices of the function $P(x,\theta,\varphi)$.

\section{Light Cone Black Holes}\label{sec:LCBH}

Now we are ready to study the different realisations of the general transformation of Minkowski spacetime and its radial MCKF to a target space time
where the conformal Killing horizons become proper Killing horizons. This transformation is expressed concisely in equation \eqref{22}.   

\subsection{Conformal compactification}\label{sec:LCBHc}

The causal structure of a generic spacetime $g_{ab}$ is easily readable once the Carter-Penrose diagram for $g_{ab}$ is found. In our case the procedure to find it is straightforward. Indeed, we already know \cite{wald2010general} that the coordinate transformation 
\be
\label{eq:einsteinuniverse}
\begin{split}
T+R &=2 \,\arctan \left(\frac{v}{\rH}\right) \\
T-R &= 2 \,\arctan \left(\frac{u}{\rH}\right) 
\end{split}
\ee
conformally maps the Minkowski metric $\eta_{ab}$  to the static Einstein Universe metric $g_{ab}^{\va EU}$. Explicitly one has
\be
\begin{split}
g_{ab}^{\va EU} &= \Omega^2_{\Mi}\, \eta_{ab}\\
-dT^2+dR^2 + \sin^2 R \,d\Omega^2 &= \Omega^2_{\Mi} (-dt^2+dr^2+r^2 d\Omega^2)
\end{split}
\ee
with
\be
\Omega^2_{\Mi} = \frac{4\rH^2}{(v^2+\rH^2)(u^2+\rH^2)}\,.
\ee
From the transformation \eqref{eq:einsteinuniverse} one can see that the Minkwoski spacetime covers only a portion of the Einstein's Universe spacetime; the boundary of which represents conformal infinity. Such portion gives the Carter-Penrose diagram of $(\R^4, \eta_{ab})$.

Any conformally flat spacetime $g_{ab}= \omega^2\,\eta_{ab}$ will be conformally mapped to the Einstein Universe by the same coordinate transformation:
\be
g_{ab}^{\va EU} = \Omega^2_{\Mi}\, \eta_{ab} =  \frac{\Omega^2_{\Mi}}{\omega^2}\,g_{ab}\equiv\Omega^2 \,g_{ab}\,.
\ee
Using \eqref{eq:omega}, the conformal factor $\Omega$ mapping the generic metric \eqref{22} to the Einstein's universe is found to be
\be\label{eq:omegaEUs}
\Omega^2 =\frac{\rH^2}{x_0^2} \frac{4\,r^2}{(v^2+\rH^2)(u^2+\rH^2)} \, P^2(x,\theta,\varphi)=\frac{\rH^2}{\rO^2-\rH^2} \frac{(v-u)^2}{(v^2+\rH^2)(u^2+\rH^2)} \, P^2(x,\theta,\varphi)\,.
\ee
As for Minkowski, the vanishing of $\Omega$ defines conformal infinity. Let us now analyse different intersting choices of the function $P(x,\theta,\varphi)$.

\subsection{The Bertotti-Robinson realisation}\label{BBRR}

\begin{figure}[t]
\center
\includegraphics[width=.4\textwidth]{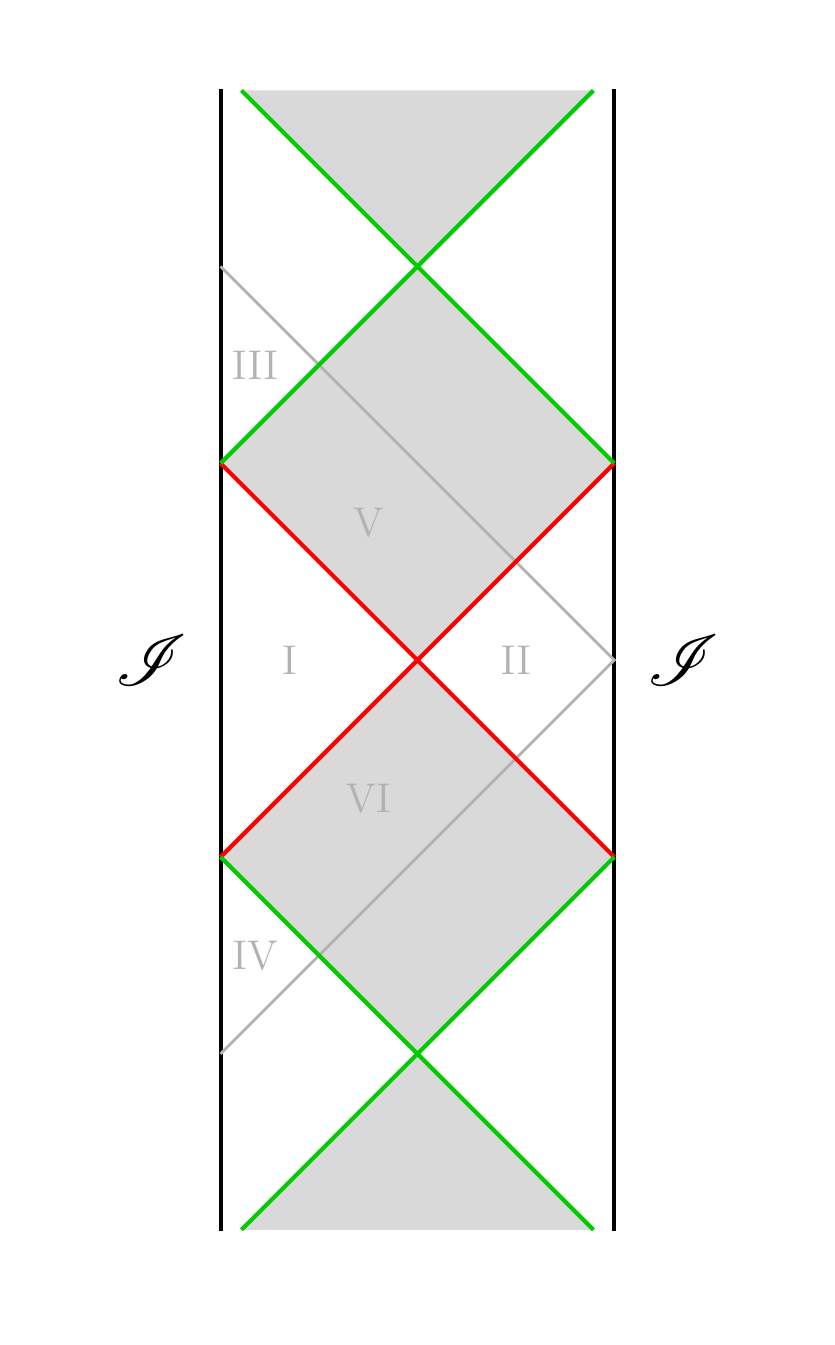} 
\caption{The causal structure of the Bertotti-Robinson spacetime. The Killing field is spacelike in the shaded regions and timelike elsewhere. The boundary is timelike and no singularities are present.  Grey lines and roman numbers show how the 6 regions of Minkowski spacetime, see Figure \ref{fig:Penrose_CD2}, are conformally mapped into the bulk of the target spacetime.}
\label{fig:Penrose_Bertotti}
\end{figure}

A compelling realisation is the simplest one: $P(x,\theta,\varphi) = 1$. From equation \eqref{eq:omega}, one can see that this spacetime is simply found dividing the Minkowski metric by $r^2$. This operation works in arbitrary dimensions, see Section \ref{nd}. 
The new metric is
\be\label{BR}
ds^2= -\frac{x^2-x_{\va BH}^2}{\xo^2}\,d\tau^2+\frac{\xo^2}{x^2-x_{\va BH}^2} \, dx^2+x_{\va 0}^2 \,d\Omega^2\,.
\ee
The Ricci and the Kretschmann scalars come out to be
\be\begin{split}
\mathbf{R}&=0 \\
\mathbf{R}_{abcd}\mathbf{R}^{abcd}&=\frac{8}{\xo^4}\,.
\end{split}\ee
Since the metric is diagonal, we can easily define a tetrad $\mathbf{e}_a{}^I$ as
\be\label{eq:tetrad}
\begin{split}
\mathbf e_\mu^0 \,dx^\mu &= \sqrt{-g_{\tau\tau}} \,dt\\
\mathbf e_\mu^1 \,dx^\mu&= \sqrt{g_{xx}} \,dr\\
\mathbf e_\mu^2\,dx^\mu &= \sqrt{g_{\theta\theta}} \,d\theta\\
\mathbf e_\mu^3 \,dx^\mu&= \sqrt{g_{\varphi\varphi}} \,d\varphi\,.
\end{split}
\ee
In this tetrad, the Einstein tensor is diagonal and given by
\be\label{eq:ene}
\mathbf{G}_{IJ}=\mathbf{G}_{ab}\,\mathbf e^a_I \,\mathbf e^b_J=\frac1{\xo^2} {\rm diag}(1,-1,1,1)\,.
\ee
The metric \eqref{BR} is a solution of Einstein-Maxwell equations for a  vector potential given by
\be\label{eBR}
\mathbf A=\frac{x}{x_{\va 0}} dt\,,
\ee
from which the electromagnetic tensor 
\be
\mathbf F=d \mathbf A=\frac{1}{x_{\va 0}} dx\wedge dt\,
\ee
follows.
The solution is static and spherically symmetric with a constant radial electric field whose flux defines the charge of the spacetime
\be
Q=\frac{1}{8\pi} \int_S \mathbf{\epsilon}_{abcd} \mathbf F^{cd}=x_{\va 0}\,.
\ee
The energy-momentum tensor satisfies the weak, strong and dominant  energy conditions. 
The spacetime is topologically $AdS_2\times S^2$. Such solution is known in the literature as the Bertotti-Robinson spacetime \cite{Bertotti:1959pf, Robinson:1959ev}. Its Carter-Penrose diagram is depicted in Figure \ref{fig:Penrose_Bertotti}.
The geometry is everywhere regular. 
There are no singularities, despite the presence of trapped surfaces and the fact that the usual energy conditions are satisfied. Singularity theorems are avoided due to the fact that 
the spacetime is not globally hyperbolic, and the generic null geodesic congruence condition \footnote{Null geodesics violating the null generic geodesic condition are those generating $\sI^{\pm}$ in Minkowski spacetime, which now pass through the bulk of the Bertotti-Robinson solution.} is not satisfied (see \cite{wald2010general} for details).
%


\subsubsection{Close to extremal Reissner-Nordstrom near-horizon geometry}

The Bertotti-Robinson solution corresponds to the near horizon geometry of a Reissner-Nordstrom (RN) black hole close to extremality---see for instance \cite{fabbri2005modeling}.
This fact in turn provides a simple interpretation to the laws of light cone mechanics  \cite{DeLorenzo:2017tgx} in terms of the standard laws of black hole thermodynamics. The RN metric for a black hole of mass $M$ and electric charge $Q$ is given by
\be
\begin{split}
ds^2=& -\left(1-\frac{2M}{r}+\frac{Q^2}{r^2}\right) dt^2+\left(1-\frac{2M}{r}+\frac{Q^2}{r^2}\right)^{-1} dr^2+r^2 d\Omega^2 \\
=&-\frac{(r-r_+)(r-r_-)}{r^2} dt^2+ \frac{r^2}{(r-r_+)(r-r_-)}dr^2+r^2 d\Omega^2, 
\end{split}
\ee
and the associated electromagnetic field by
\be
\mathbf A=-\frac{Q}{r} dt,
\ee
where $r_{\pm}=M\pm\sqrt{M^2-Q^2}$. The near extremal case corresponds to  $M=Q+\delta M$ with $\delta M^2 \ll Q^2$, for which
the near horizon metric and electromagnetic field is obtained by expanding in the new coordinate $x$ defined by $r=x_{\va 0}+x$.
The leading order gives the metric \eqref{BR} and electromagnetic field \eqref{eBR} with $x_{\va 0}=Q$ and
$\xBH=\sqrt{2\,Q\,\delta m}$. Therefore, we can relate the physical parameter of the RN solution to the parameters of the MCKF in flat spacetime via
\be
Q^2={\rO^2-\rH^2} \qquad {\rm and} \qquad \delta m = \frac{2 \rH^4}{(\rO^2-\rH^2)^{\frac32}}\,.
\ee
The RN Hawking temperature $T_{\va RN}$ is given by the standard formula
\be
T_{\va RN}=\frac{\kappa_{\va RN}}{2\pi}=\frac{\rH}{\pi (\rO^2-\rH^2)}
\ee
in agreement with \eqref{pepino}. 
This shows that the limit $\rH\to 0$ corresponds exactly to the extremal limit of the RN solution. On the RN side the temperature goes to zero and the bifurcating sphere goes away to infinity. On the Minkowski side, the radius of the bifurcating sphere $\rH$ shrinks to zero and the conformal Killing horizon becomes the light cone of 
a single event \cite{DeLorenzo:2017tgx}. 

\subsubsection{Near horizon symmetry is radial conformal symmetry in $\M^4$}\label{nhs}

Consider the $2$-dimensional metric
\be\label{westa}
ds^2= -\frac{x^2-x_{\va BH}^2}{\xo^2}\,d\tau^2+\frac{\xo^2}{x^2-x_{\va BH}^2} \, dx^2.
\ee
This metric is locally $AdS_2$ and therefore its isometry group is $SL(2,\R)$. There are therefore three independent Killing fields which are also generators of isometries of \eqref{BR}. Via the conformal transformation that relates this spacetime to Minkowski spacetime we infer that the $SL(2,\R)$ generators should correspond to conformal Killing vectors in Minkowski spacetime. 

We observe that the vector field
\be
v_1=\tau \partial_\tau-x\partial_x
\ee
is the generator of an infinitesimal diffeomorphism sending the metric \eqref{westa} to a new one where the $\xBH\to \xBH (1-\alpha)$, with $\alpha$  the infinitesimal parameter of the transformation. Indeed the previous is a Killing field for the metric \eqref{westa} with $\xBH=0$. 

The following coordinate transformation
\be\label{cocorico}
\begin{split}
z=&\frac{\xo}{\xBH}\exp\left(-\frac{\xBH}{\xo^2}\tau\right)\sqrt{x^2-\xBH^2} \\
t=&\frac{x \xo\exp\left(\frac{\xBH}{\xo^2}\tau\right)}{\sqrt{x^2-\xBH^2} }=\frac{\xo^2}{\xBH}\frac{x}{z}
\end{split}
\ee
transforms the metric \eqref{westa} into
\be
ds^2= -\frac{z^2}{\xo^2}\,dt^2+\frac{\xo^2}{z^2} \, dz^2
\ee
Three independent Killing fields are known in the previous coordinates, namely
\be\begin{split}
v_1=& {t}\partial_t-{z}\partial_z=\frac{(v^2-\rH^2)}{2\rH} \partial_v +\frac{(u^2-\rH^2)}{2\rH} \partial_u\\
v_2=&\xo \partial_t=\frac{(v+\rH)^2}{2\rH} \partial_v +\frac{(u+\rH)^2}{2\rH} \partial_u \\
v_3=&\left(\frac{\xo^2}{2z^2}+\frac{t^2}{2\xo^2}\right)\xo\partial_t-\frac{tz}{\xo}\partial_z
=\frac{(v-\rH)^2}{4\rH} \partial_v+\frac{(u-\rH)^2}{4\rH} \partial_u,
\end{split}\ee
where on the right we have used the coordinate transformations back to $uv$-Minkowski null coordinates using \eqref{eq:CooTra} combined with \eqref{cocorico}. 
We recognise the Killing field of departure \be
\xi= \frac{2\rH}{\rO^2-\rH^2} v_1\ee
The commutator algebra is
\be\begin{split}
\left[v_2,v_1\right]=&v_2 \\
\left[v_2,v_3\right]=&v_1 \\
\left[v_1,v_3\right]=&v_3
\end{split}\ee
which corresponds to the $SL(2,\R)$ Lie algebra \eqref{lieSL} for the generators $T_0=(v_3+v_2)/\sqrt{2}$, $T_2=(v_3-v_2)/\sqrt{2}$, and $T_1=-v_1$. Notice also that in $uv$-coordinates the radial conformal symmetry group $SL(2,\R)=SL(2,\R)_{in}\times SL(2,\R)_{out}$ as in the $AdS_2$ symmetry of near horizon geometry.
\subsection{De Sitter realisation}\label{sub:desitter}

Another interesting case arises by sending a constant $x=x_*$ Minkowski conformal Killing observer to infinity via the choice of $P(x,\theta,\phi)$ in \eqref{22}. From the discussion under equation \eqref{eq:omegaEUs}, this is achieved by choosing a function $P(x,\theta,\phi)$ which vanishes at $x=x_*$. The simplest choice admitting a regular differential structure at infinity is
\be\label{chochi}
P(x) = \frac{x_*-x}{\xo}\,.
\ee
The corresponding metric is therefore
\be
ds^2=\frac{\xo^2}{(x_*-x)^2} \left(-\frac{x^2-x_{\va BH}^2}{\xo^2}\,d\tau^2+\frac{\xo^2}{x^2-x_{\va BH}^2} \, dx^2+x_{\va 0}^2 \,d\Omega^2\right)\,.
\ee
In the new coordinate
\be\label{X}
X^2 = \frac{\xo^4}{(x_*-x)^2}
\ee
the metric takes the simple form
\be\label{ds}
ds^2=-F(X)\,d\tau^2+\frac{1}{F(X)} \, dX^2+X^2 \,d\Omega^2
\ee
with
\be
F(X) = 1-\frac{2\,x_* }{\xo^2}\,X+\frac{(x_*^2-\xBH^2)}{\xo^4} \,X^2\,.
\ee
The surface $x_*$ sent to infinity corresponds now to $X\to +\infty$.
The Ricci scalar for this new solution turns out to be
\be
\mathbf{R}=-12 \left(\frac{x_*^2-\xBH^2}{\xo^4}+\frac{x_*}{\xo^2\,X}\right)
\ee
which tends to a constant as $X \to +\infty$
\be\label{eq:lambda}
4\,\Lambda = \lim_{X\to +\infty} \mathbf{R}=-12\,\frac{x_*^2-\xBH^2}{\xo^4}\,.
\ee
Such constant is positive if $x_*$ is chosen in between the inner and outer horizons and negative elsewhere.
Moreover, for $x_*\neq 0$, $\mathbf{R}$ diverges as $X$ approaches zero
\be
\mathbf{R}|_{X\to 0} = -12\, \frac{x_*}{\xo^2\,X}+ O(1)\,.
\ee
As in the previous case, we can define a diagonal tetrad as in \eqref{eq:tetrad}, which gives the diagonal Einstein's tensor
\be\label{eq:eneAdS}
\begin{split}
\mathbf G_{00}&=-3\,\frac{x_*^2-\xBH^2}{\xo^4}+4\,\frac{x_*}{\xo^2\,X}\\
\mathbf G_{11}&=3\,\frac{x_*^2-\xBH^2}{\xo^4}-4\,\frac{x_*}{\xo^2\,X}\\
\mathbf G_{22}&=3\,\frac{x_*^2-\xBH^2}{\xo^4}-2\,\frac{x_*}{\xo^2\,X}\\
\mathbf G_{33}&=3\,\frac{x_*^2-\xBH^2}{\xo^4}-2\,\frac{x_*}{\xo^2\,X}\,.
\end{split}
\ee
The metric can therefore be interpreted as a solution to the Einstein's equation with a cosmological constant $\Lambda$ as in  \eqref{eq:lambda} and a energy-momentum tensor given by
\be
\mathbf T_{IJ} = \frac{2\,x_*}{\xo^2\,X} \,\text{diag}(2,-2,-1,-1)\,.
\ee
The global as well as the local nature of these spacetimes depends on the explicit value of $x_*$.

\begin{figure}[t]
\center
\includegraphics[width=.6\textwidth]{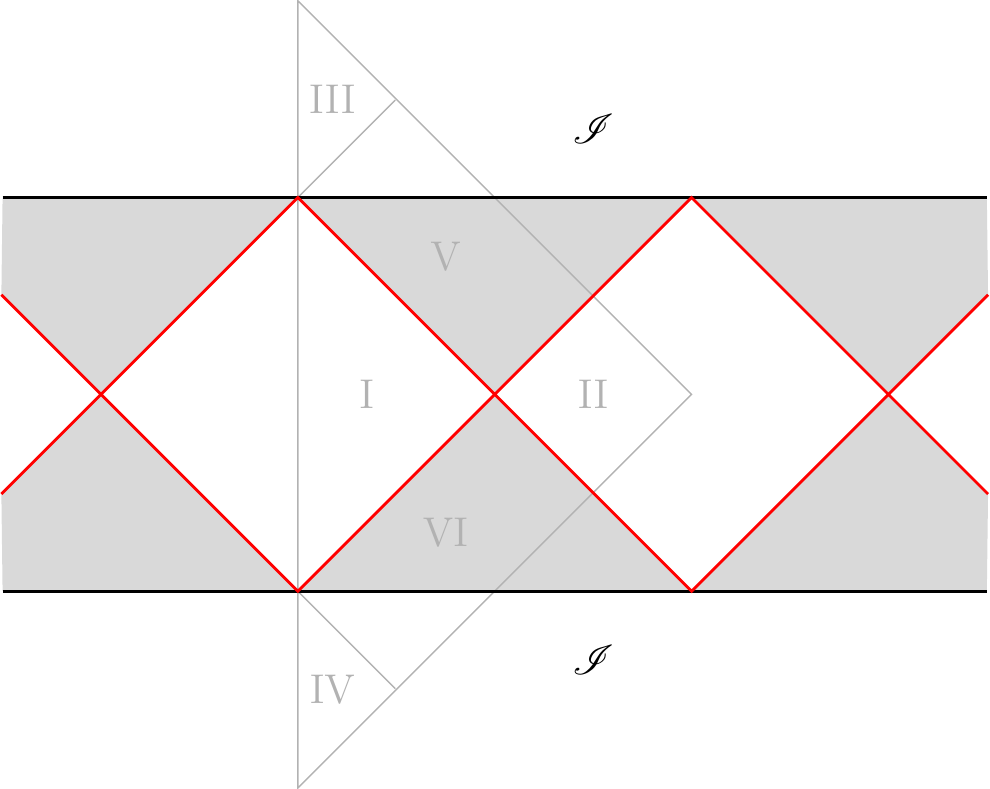}
\caption{The causal structure of the {de Sitter} spacetime obtained from Minkowski by choosing $x_*=0$ in  \eqref{chochi} including the MCKF. The Killing field is spacelike in the shaded regions and timelike elsewhere.  Grey lines and numbers show how the 6 regions of Minkowski spacetime depicted in Figure \ref{fig:Penrose_CD2} are conformally mapped into the bulk of the target spacetime. The light cones in flat spacetime are mapped to intersecting cosmological horizons.}
\label{fig:Penrose_dS}
\end{figure}

For $x_*=0$, $\mathbf T_{ab}=0$, the Ricci scalar is non-diverging, and the Einstein tensor \eqref{eq:eneAdS} corresponds to that of a positive cosmological constant term $\Lambda \,g_{ab}$ with
\be
\Lambda=3\frac{\xBH^2}{\xo^4}\,.
\ee
For this choice of $x_*$, the metric \eqref{ds} is manifestly that of { de Sitter} spacetime in terms of static coordinates. The bifurcating Killing horizon corresponds to the union of a past and future cosmological horizons intersecting at the bifurcating sphere, as shown in the Carter-Penrose diagram in Figure \ref{fig:Penrose_dS}. 

\subsection{Asymptotically (Anti){-de Sitter} realisations}\label{sub:asyAdS}

If $x_*< 0$ one has a positive cosmological constant for $|x_*|>|\xBH|$ and a negative one for $|x_*|<|\xBH|$  with a $\mathbf T_{ab}$ violating all the standard energy conditions. In these cases the metric is asymptotically $dS$ and $AdS$ respectively. In the $AdS$ case the decay rate to the asymptotic geometry is slower than the one imposed by the standard reflecting boundary conditions \cite{Ashtekar:1999jx}. This implies that a well-defined notion of conserved charges at infinity is not possible. The spacetime is therefore in this sense weakly asymptotically $AdS$. 

\begin{figure}[t]
\centering
\hspace{-1.2cm}
\subfigure[Weakly asymptotically AdS.]{
   \includegraphics[width=.4\textwidth]{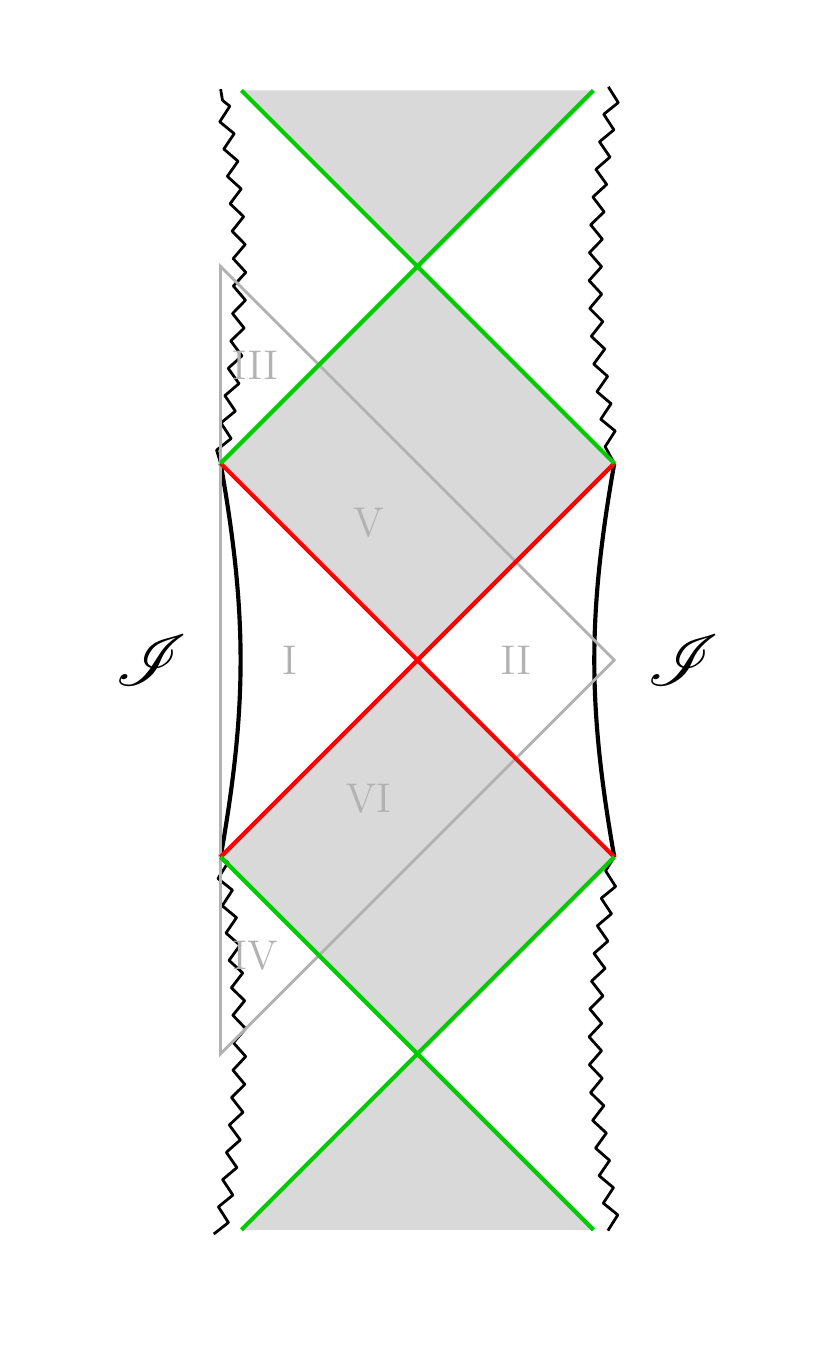}}\hspace{2cm}
\subfigure[Asymptotically dS.]{\raisebox{1cm}{
   \includegraphics[width=.26\textwidth]{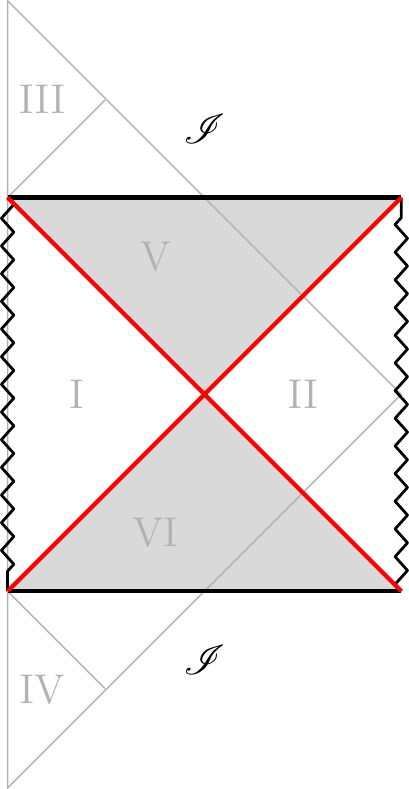}}}
\caption{The causal structure of the different spacetime realisations when an observer is sent to infinity. The two spherical dimensions are suppressed, so that each point represents a sphere. The Killing field is spacelike in the shaded regions and timelike elsewhere. Grey lines and numbers show how the 6 regions of Minkowski spacetime depicted in Figure \ref{fig:Penrose_CD2} are conformally mapped into the bulk of the target spacetime.}
\label{fig:Penrose_(A)dS}
\end{figure}

If $x_*>0$ instead, one has a positive asymptotic cosmological constant for $x_*<\xBH$ and a negative one for $x_*>\xBH$. Now $\mathbf T_{ab}$ satisfies the {\em weak} ($\rho\ge 0, \rho+p_i\ge 0$) and the {\em dominant} ($\rho\ge |p_i|$) energy condition but not the {\em strong} one ($\rho+p_i\ge 0, \rho+\sum_i p_i\ge 0$). For negative asymptotic cosmological constant, there are black hole regions, plus inner and outer Killing horizons. The boundary is again weakly $AdS$. The new feature with respect to the Bertotti-Robinson realisation is the appearance of a time like curvature singularity at $X=0$. There is no black hole region in the $dS$ realisation, while the time like curvature singularities at $X=0$ remain.
The Carter-Penrose diagrams corresponding to these cases are shown in Figure \ref{fig:Penrose_(A)dS}. 

\subsection{An asymptotically flat realisation}\label{originalone}
The inner horizons of the Bertotti-Robinson realisation of Section \ref{BBRR} can become the boundary of the spacetime via a particular choice of $P(x,\theta,\varphi)$ in \eqref{22}.
Such realisation was already evoked in  \cite{DeLorenzo:2017tgx} to illustrate some aspects of the light cone thermodynamical laws, but its global structure was not properly studied.
In this case the metric is
\be
g_{ab} = \frac{16\,\rH^4}{(u-\rH)^2(v+\rH)^2}\,\eta_{ab}\,.
\ee
In Appendix~\ref{CooTra}, see equations (\ref{ji1}) and (\ref{ji2}), a coordinate transformation is found such that the metric is
\be\label{aaaa}
ds^2=\frac{1}{\Delta}\left(-(1-2\,z)\,d\bar{\tau}^2+\frac{1}{1-2\,z} \,dz^2+z^2\,d\Omega^2\right)\,,
\ee
where the new coordinates are dimensionless and $\Delta = \frac{4\rH^2}{\xo^4}$.
In these coordinates the horizon is located at $z=1/2$. Moreover, $z$ is positive and greater than $1/2$ outside, and decreases to zero at the Minkowskian $i^0$ and origin. Inside the horizon, on the other hand, $z$ increases from $z=1/2$ to $z\to \infty$. The latter corresponds to the inner horizon which, now we show, is at infinity in the target spacetime. From the discussion of Section~\ref{sec:LCBHc} we can draw its Carter-Penrose diagram and study the properties of the corresponding conformal boundary.
The conformal factor $\Omega^2$ mapping \eqref{aaaa} to the Einstein's Universe $g_{ab} = \Omega^2\,g_{ab}^{\va EU}$ is given by
\be\begin{split}
\Omega^2 &=\frac{1}{\rO^2-\rH^2} \frac{(u-\rH)^2(v+\rH)^2}{4\,(u^2+\rH^2)(v^2+\rH^2)}\\
&=\frac{1}{\rO^2-\rH^2} \frac{(1-\sin U)(1+\sin V)}{4}\\
&=\frac{1}{\rO^2-\rH^2} \left(\frac{\cos T+\sin R}{2}\right)^2
\end{split}
\ee 
where $U=T-R$ and $V=T+R$ are the Einstein's Universe null coordinates. Conformal infinity $\scri$ is given by the condition $\Omega=0$, and therefore
\be
\scri:\,\qquad T-R=U=\frac{\pi}{2} \qquad{\rm and} \qquad T+R=V=-\frac{\pi}2\,.
\ee
which is equivalent to $u=\rH$ and $v=-\rH$, or simply $z\to \infty$. The boundary is made of two constant $U$ or $V$ surfaces, being therefore null.

\begin{figure}[t]
\center
\includegraphics[width=.2\textwidth]{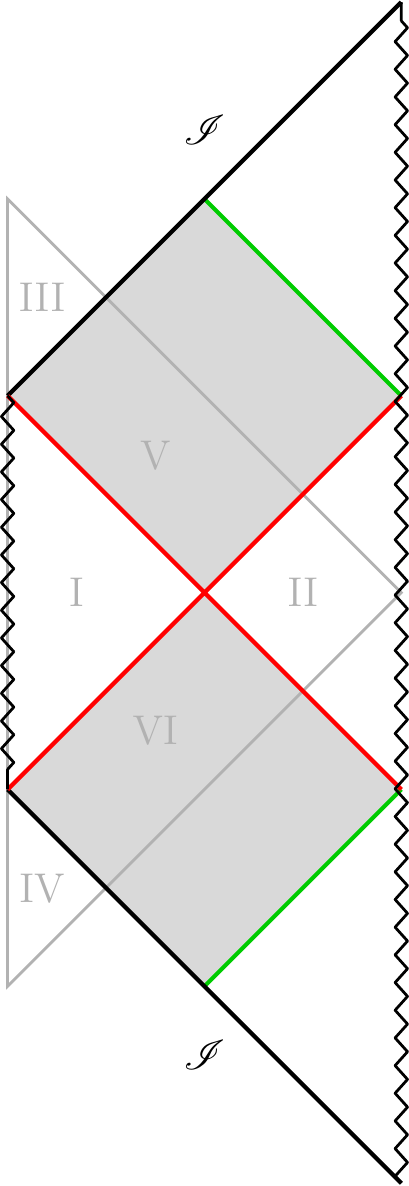} 
\caption{The causal structure of the conformally flat realisation. The two spherical dimensions are suppressed, so that each point represent a sphere.  The grey lines and numbers show how the 6 regions of Minkowski spacetime depicted in Figure \ref{fig:Penrose_CD2} are conformally mapped into the bulk of the target spacetime.}
\label{fig:Penrose_butterfly}
\end{figure}

The gradient of the conformal factor is found to be
\be
\big(\tilde\nabla_\mu \omega\big)\,dx^\mu =-\frac{\sin T}{2 \rH}\, dT+\frac{\cos R}{2 \rH}\,dR,
\ee
which is non-zero at $\scri$. The Ricci tensor in the diagonal tetrad is
\be
\mathbf{R}_{IJ} = \mathbf{R}_{ab}\,\mathbf{e}^a_I \mathbf{e}^b_J = \frac{2\Delta}{z}{\rm diag}(-1,1,2,2)
\ee
which vanishes in a neighbourhood of $\scri$, i.e. for $z\to \infty$. Our spacetime fulfills all the conditions of the definition of conformally flatness \cite{frauendiener2004conformal}.
Finally, the Ricci scalar is
\be
\mathbf{R} = \frac{12 \, \Delta}{z}
\ee
showing a curvature singularity at $z=0$. The resulting causal structure is shown in Figure \ref{fig:Penrose_butterfly}.
For this last case we made explicit the analysis for the construction of the Carter-Penrose diagram. The very same strategy is used for the previous cases as well, but, for brevity, we preferred not to present it explicitly.

\section{Arbitrary Dimensions}\label{nd}

The construction of Section \ref{BBRR} works in arbitrary dimensions. Namely, the simple rescaling of the $n$-dimensional Minkowski metric by the factor $\omega^2=\xo^2/r^2$, i.e. defining $g_{ab}=\omega^2 \eta_{ab}$, turns the conformal Killing field \eqref{killingg} into a Killing field of the target spacetime. This is clear from the fact that
\be\begin{split}
 \nabla_a\xi^a=&\frac{1}{\sqrt{|g|}}\partial_{\mu} \left(\sqrt{|g|} \xi^\mu\right) \\
 =&\frac{r^n}{\sqrt{|\eta|}}\partial_{\mu} \left( \frac{\sqrt{|\eta|}}{r^n} \xi^\mu\right)\\
 =&{r^2}\partial_{r} \left( \frac{\xi^r}{r^2}  \right)+\partial_{t} \xi^t =0,
 \end{split} \ee
 where $\eta$ is the determinant of the $n$-dimensional Minkowski metric, and where we have used that,  in spherical coordinates, $\sqrt{|\eta|}=r^{n-2} \sin({\theta_1})\cdots\sin(\theta_{n-3})$ for $n>2$ and $\sqrt{|\eta|}=1$ for $n=2$. The vanishing of the last line follows from the direct substitution of the components of $\xi^a$ given in \eqref{killingg}. Moreover, this is also evident from the direct application of the coordinate transformation \eqref{eq:CooTra} to $({\xo^2}/{r^2}) ds_{\va M}^2$ which yields the Bertotti-Robinson spacetime in arbitrary dimensions, namely
\be\label{hihin}
ds^2= -\frac{x^2-x_{\va BH}^2}{\xo^2}\,d\tau^2+\frac{\xo^2}{x^2-x_{\va BH}^2} \, dx^2+x_{\va 0}^2 \,dS_{n-2}^2\,,
\ee
where $dS_{n-2}^2$ denotes the metric of the $(n-2)$-dimensional unit sphere. The other realisations can also be constructed along the same lines.

\subsection{The Jackiw-Teitelboim realisation}

An especially interesting case is the case $d=2$ where the natural conformal mapping corresponding to the two-dimensional version of (\ref{22}) with $P(x)=1$ produces
a well studied black hole solution of dilaton gravity know as the Jackiw-Teitelboim model \cite{jackiw1984quantum}.

\section{Conclusions}
A mathematical analogy of the laws of black hole thermodynamics  for light cones in Minkowski spacetime was found in \cite{DeLorenzo:2017tgx}. This was possible by observing that intersecting light cones are bifurcating conformal Killing horizons for the most general radial conformal Killing field $\xi$. The causal behavior of $\xi$ closely resembles the one of the stationarity Killing field of a Reissner-Nordstrom black hole---see Figure \ref{fig:penrose}. Such conformal stationarity is the heart of the validity of the four  laws of light cone thermodynamics introduced in \cite{DeLorenzo:2017tgx}. However, the quantities appearing in these laws have no direct physical meaning in flat space. Nevertheless,  as they are all conformally invariant, they acquire their standard geometric, as well as physical, meaning in conformally flat spacetimes where the conformal Killing field becomes a genuine Killing field. 

In this work we have studied the properties of representative spacetimes satisfying this condition. The most interesting case is the simplest one. It turns out that the conformal Killing horizon structure of light cones in Minkowski spacetime is the conformal partner of the Killing horizon structure of the Bertotti-Robinson spacetime. The latter is known to encode  the near horizon geometry of close-to-extremal and extremal charged black holes. The reason behind this fact is the vanishing of the Weyl tensor at the horizon of an extremal Reissner-Nordstrom black hole \cite{Chandrasekhar:1985kt} which makes the near horizon geometry conformally flat.

This result completely clarifies the nature of light cone thermodynamics as introduced in \cite{DeLorenzo:2017tgx}: it can now be seen as arising from a conformal transformation of the standard laws of black hole thermodynamics on a suitable black hole spacetime. This strengthens our initial claim that light cones in Minkowski spacetime encode, in a suitable sense, the main properties of black hole horizons, thus providing a simple analogue of black holes in a spacetime with trivial curvature. The analogy is more strict and direct than the one usually considered between near horizon black hole geometry and Rindler spacetime.  The black hole and Rindler horizons, indeed, have different topologies, being respectively $ S^2 \times \mathbb{R}$ and $\mathbb{R}^2\times \mathbb{R}$. Additionally, the Rindler wedge cannot be seen as the region outside the horizon, since it lies itself in the domain of dependence of the latter. This in turn implies that no finite energy flux can escape the Rindler horizon, and no notion of asymptotic observer can be defined. These difficulties are not present in the light cone case studied here and in \cite{DeLorenzo:2017tgx}. As for black holes, the light cone topology is $S^2\times \mathbb{R}$, and energy can be sent to infinity without crossing the horizon from the complement of the diamond, Region II in Figure \ref{fig:Penrose_CD2}, which therefore plays the role of the outside region. The analogy is indeed so strict that for conformally invariant matter the light cone structure is indistinguishable from the near horizon geometry of a close-to-extremal Reissner-Nordstrom black hole. Other interesting conformally flat spacetimes where the conformal Killing horizon structure becomes a proper Killing one have been presented. 

One would be tempted to say that an analog to a spinning near extremal black hole in dimension four can also be obtained in the present way. However, this is not so since the extremal Kerr near horizon geometry is not conformally flat \cite{Chandrasekhar:1985kt}. Analogue of spinning black holes require a different approach. Nevertheless, the argument fails in three dimensions where spinning black holes are represented by the Banados-Teitelboim-Zanelli (BTZ) solutions \cite{Banados:1992wn}. The BTZ black holes are indeed conformally flat and should have a representation in terms of light surfaces in Minkowski spacetime. It can be seen however that the Killing generator of the BTZ horizon cannot correspond to our radial MCKF.  The correct mapping to flat spacetimes will have to involve the topological identifications \cite{Banados:1992gq}, suitably translated to Minkowski, that are necessary in $AdS_3$ to obtain the BTZ geometry.

Our analysis emphasises the conformally flat nature of the near extremal near horizon approximation of RN spacetimes, as well as the obvious case of De Sitter spacetimes. 
This allows for a simple analysis of the renormalization of the energy momentum tensor for conformal fields and the immediate computation of $\langle T_{ab}\rangle$ in these cases \cite{Wald:1978ce}.
Along these lines, our analysis could be used to provide a simple interpretation to the computation of (logarithmic corrections to) entanglement entropy for bifurcate Killing horizons from the perspective of quantum field theory on Minkowski spacetimes.    

The near horizon geometry symmetry structure of near extremal and extremal Reissner-Nordstrom black holes is present in flat spacetimes and is associated to the algebra of radial conformal Killing fields of Minkowski. The basic ingredients of the extremal black holes  CFT correspondence conjecture \cite{Guica:2008mu, Compere:2012jk} might be available here. Notice however that in the Kerr-CFT correspondence the conformal symmetry concerns the $\varphi-t$ ``plane'' (for RN it works by adding a Kaluza-Klein dimension $y$ for the electromagnetic unification and the conformal symmetry occurs one dimension up in the $y-t$ ``plane'') \cite{Hartman:2008pb}.  Another approach to entropy based on the appearance of a conformal symmetry in the $r-t$ ``plane'' is the one studied by Carlip  \cite{Carlip:1998wz}. It would be nice to investigate the possibility of a formulation of such ideas entirely in the context of a flat background. We leave these investigations for the future.

\section{Acknowledgments}
We would like to thank Abhay Ashtekar for useful insights on the asymptotic structure of $AdS$ spacetimes. We acknowledge support from the   OCEVU   Labex   (ANR-11-LABX-0060)   and   the
A*MIDEX  project  (ANR-11-IDEX-0001-02)  funded  by the ``Investissements d'Avenir'' French government program managed by the ANR. This work was supported in part by the Eberly Research Funds of Penn State and by the Pittsburgh Foundation UN2017-92945 as well as the NSF-NSF Investigations PHY-{1806356}.

\begin{appendix}

\section{Coordinate Transformations}\label{CooTra}
In \cite{DeLorenzo:2017tgx}, six different coordinate transformations $(t,r,\theta,\varphi) \to (\tau,\rho,\theta,\varphi)$ for each and every of the six regions Minkowski spacetime is divided into by the radial MCKF was presented. The transformation was built so that the radial MCKF simplified  to 
\be\label{eq:killtauapp}
\xi^\mu \frac{\partial}{\partial x^\mu} = \frac{\partial}{\partial \tau}\,.
\ee
Here we observe that the six transformations can actually be grouped in one single transformation given by
\be
\begin{split}
\tau &= \frac{\rO^2-\rH^2}{4\rH}\log\frac{(u-\rH)(v-\rH)}{(u+\rH)(v+\rH)}\\
\rho &= \frac{\rO^2-\rH^2}{4\rH}\log\frac{(u+\rH)(v-\rH)}{(u-\rH)(v+\rH)}\,.
\end{split}
\ee
The coordinate $\tau$ is the same used in the main text, equation \eqref{eq:CooTra}.
Defining as in \cite{DeLorenzo:2017tgx}
\be
\begin{split}
\Delta &= \frac{4\rH^2}{(\rO^2-\rH^2)^2}\\
a &= \frac{1}{\rO^2-\rH^2}
\end{split}
\ee Minkowski metric becomes
\be
ds_\Mi^2=\left(\frac{\Delta/2a}{\cosh(\sqrt{\D}\, \tau)+\cosh(\sqrt{\D} \,\rho)}\right)^2\big(-d \tau^2+d\rho^2+\Delta^{-1}\sinh^2(\sqrt{\D}\,\rho)\,d\Omega^2\big)\,.
\ee
The transformation is valid everywhere using the standard definition of the logarithm of a negative number, namely
\be
\log (-x) = i\,\pi + \log(x) \qquad x>0\,.
\ee
We can solve equation \eqref{eq:conftokill} in these coordinates finding
\be
\omega^2 = \left(\frac{\cosh(\sqrt{\D}\, \tau)+\cosh(\sqrt{\D} \,\rho)}{\D/2a}\frac{1}{G_{\rho}(\rho)}\right)^2\,,
\ee
which 
in terms of the Minkowskian double-null coordinates $(u,v)$ is
\be\label{eq:omega2}
\omega^2(u,v) = \frac{4r_{\va H}^4}{(u^2-\rH^2)(v^2-\rH^2)} \frac{1}{G^2(\rho,\theta,\phi)}\,.
\ee
A conformally flat metric $g_{ab}$ such that the radial MCKF becomes a Killing field can therefore also be written as
\be\label{metrho}
ds^2=\frac{1}{G^2_\rho(\rho,\theta,\phi)}(-d \tau^2+d\rho^2+\Delta^{-1}\sinh^2(\sqrt{\D}\rho)\,d\Omega^2)\,,
\ee
Choosing the function $G_{\rho}$ to be a normalisation constant given by
\be
G_{\rho}= \frac{\rO^2-\rH^2}{2\rH^2}
\ee
one finds 
\be
\omega_{\va FRW} = \frac{\rO^2-\rH^2}{\sqrt{(u^2-\rH^2)(v^2-\rH^2)}} 
\ee
The choice
\be\label{PBH}
G_{\rho}(\rho)=1/4\,e^{-2\sqrt{\D}\,\rho}\,,
\ee
instead, gives
\be
\omega_{\va BH}=\frac{4r^2_{\va H}}{(u-r_{\va H})(v+r_{\va H})}.
\ee
The above two conformal factors were found in \cite{DeLorenzo:2017tgx} by separation of variables. 

Another interesting coordinates transformation is given by
\be\label{ji1}
\begin{split}
\bar{\tau} &= \sqrt{\Delta}\,\tau\\
\rho &= \frac{1}{2\sqrt{\D}}\log (1-2z)\,.
\end{split}
\ee
which implies
\be\label{ji2}
z=\frac12\left(1-\frac{(u+\rH)(v-\rH)}{(u-\rH)(v+\rH)}\right)\,.
\ee
The metric \eqref{metrho} takes 
the following Schwarzschild-like form
\be
ds^2=\frac{\Delta^{-1}}{G^2_z(z,\theta,\phi)} \left(-(1-2\,z)\,d\bar{\tau}^2+\frac{1}{1-2\,z} \,dz^2+z^2\,d\Omega^2\right)\,,
\ee
where $G_z^2$ is a new function encoding the ambiguity in the conformal transformations, and the new coordinates are dimensionless. In these coordinates the horizon is located at $z=1/2$. $z$ is positive and greater then $1/2$ outside, and decreases to zero at the Minkowskian $i^0$ and origin. Inside the horizon, on the other hand, increases from $z=1/2$ to $z\to \infty$, the latter corresponding to the inner horizon. In Section~\ref{originalone}, these coordinates are used in the simplest case $G_z=1$.

\end{appendix}

\bibliographystyle{JHEP}
\bibliography{references}

\providecommand{\href}[2]{#2}\begingroup\raggedright\begin{thebibliography}{10}

\bibitem{RCKF}
A.~Herrero and J.~A. Morales, \emph{Radial conformal motions in minkowski
  space--time}, \href{https://doi.org/10.1063/1.532903}{\emph{Journal of
  Mathematical Physics} {\bfseries 40} (1999) 3499--3508},
  [\href{https://arxiv.org/abs/http://aip.scitation.org/doi/pdf/10.1063/1.532903}{{\ttfamily
  http://aip.scitation.org/doi/pdf/10.1063/1.532903}}].

\bibitem{DeLorenzo:2017tgx}
T.~De~Lorenzo and A.~Perez, \emph{{Light Cone Thermodynamics}},
  \href{https://doi.org/10.1103/PhysRevD.97.044052}{\emph{Phys. Rev.}
  {\bfseries D97} (2018) 044052},
  [\href{https://arxiv.org/abs/1707.00479}{{\ttfamily 1707.00479}}].

\bibitem{Jacobson:1993pf}
T.~Jacobson and G.~Kang, \emph{{Conformal invariance of black hole
  temperature}},
  \href{https://doi.org/10.1088/0264-9381/10/11/002}{\emph{Class. Quant. Grav.}
  {\bfseries 10} (1993) L201--L206},
  [\href{https://arxiv.org/abs/gr-qc/9307002}{{\ttfamily gr-qc/9307002}}].

\bibitem{wald2010general}
R.~M. Wald, \emph{General relativity}.
\newblock University of Chicago press, 1984.

\bibitem{Bertotti:1959pf}
B.~Bertotti, \emph{{Uniform electromagnetic field in the theory of general
  relativity}}, \href{https://doi.org/10.1103/PhysRev.116.1331}{\emph{Phys.
  Rev.} {\bfseries 116} (1959) 1331}.

\bibitem{Robinson:1959ev}
I.~Robinson, \emph{{A Solution of the Maxwell-Einstein Equations}},
  {\emph{Bull. Acad. Pol. Sci. Ser. Sci. Math. Astron. Phys.} {\bfseries 7}
  (1959) 351--352}.

\bibitem{jackiw1984quantum}
R.~Jackiw and C.~Teitelboim, \emph{Quantum theory of gravity}, {\emph{Adam
  Hilger, Bristol} (1984) }.

\bibitem{francesco2012conformal}
P.~Di~Francesco, P.~Mathieu and D.~Senechal, \emph{{Conformal Field Theory}}.
\newblock Graduate Texts in Contemporary Physics. Springer-Verlag, New York,
  1997,
  \href{https://doi.org/10.1007/978-1-4612-2256-9}{10.1007/978-1-4612-2256-9}.

\bibitem{Martinetti:2002sz}
P.~Martinetti and C.~Rovelli, \emph{{Diamonds's temperature: Unruh effect for
  bounded trajectories and thermal time hypothesis}},
  \href{https://doi.org/10.1088/0264-9381/20/22/015}{\emph{Class. Quant. Grav.}
  {\bfseries 20} (2003) 4919--4932},
  [\href{https://arxiv.org/abs/gr-qc/0212074}{{\ttfamily gr-qc/0212074}}].

\bibitem{Martinetti:2008ja}
P.~Martinetti, \emph{{Conformal mapping of Unruh temperature}},
  \href{https://doi.org/10.1142/S0217732309030874}{\emph{Mod. Phys. Lett.}
  {\bfseries A24} (2009) 1473--1483},
  [\href{https://arxiv.org/abs/0803.1538}{{\ttfamily 0803.1538}}].

\bibitem{Bianchi:2013rya}
E.~Bianchi and A.~Satz, \emph{{Mechanical laws of the Rindler horizon}},
  \href{https://doi.org/10.1103/PhysRevD.87.124031}{\emph{Phys. Rev.}
  {\bfseries D87} (2013) 124031},
  [\href{https://arxiv.org/abs/1305.4986}{{\ttfamily 1305.4986}}].

\bibitem{fabbri2005modeling}
A.~Fabbri and J.~Navarro-Salas, \emph{Modeling black hole evaporation}.
\newblock World Scientific, 2005.

\bibitem{Ashtekar:1999jx}
A.~Ashtekar and S.~Das, \emph{{Asymptotically Anti-de Sitter space-times:
  Conserved quantities}},
  \href{https://doi.org/10.1088/0264-9381/17/2/101}{\emph{Class. Quant. Grav.}
  {\bfseries 17} (2000) L17--L30},
  [\href{https://arxiv.org/abs/hep-th/9911230}{{\ttfamily hep-th/9911230}}].

\bibitem{frauendiener2004conformal}
J.~Frauendiener, \emph{Conformal infinity}, {\emph{Living Reviews in
  Relativity} {\bfseries 7} (2004) 1}.

\bibitem{Chandrasekhar:1985kt}
S.~Chandrasekhar, \emph{{The mathematical theory of black holes}},  in
  \emph{{Oxford, UK: Clarendon (1992) 646 p.}}, 1985.

\bibitem{Banados:1992wn}
M.~Banados, C.~Teitelboim and J.~Zanelli, \emph{{The Black hole in
  three-dimensional space-time}},
  \href{https://doi.org/10.1103/PhysRevLett.69.1849}{\emph{Phys. Rev. Lett.}
  {\bfseries 69} (1992) 1849--1851},
  [\href{https://arxiv.org/abs/hep-th/9204099}{{\ttfamily hep-th/9204099}}].

\bibitem{Banados:1992gq}
M.~Banados, M.~Henneaux, C.~Teitelboim and J.~Zanelli, \emph{{Geometry of the
  (2+1) black hole}}, \href{https://doi.org/10.1103/PhysRevD.48.1506,
  10.1103/PhysRevD.88.069902}{\emph{Phys. Rev.} {\bfseries D48} (1993)
  1506--1525}, [\href{https://arxiv.org/abs/gr-qc/9302012}{{\ttfamily
  gr-qc/9302012}}].

\bibitem{Wald:1978ce}
R.~M. Wald, \emph{{Axiomatic Renormalization of the Stress Tensor of a
  Conformally Invariant Field in Conformally Flat Space-Times}},
  \href{https://doi.org/10.1016/0003-4916(78)90040-4}{\emph{Annals Phys.}
  {\bfseries 110} (1978) 472--486}.

\bibitem{Guica:2008mu}
M.~Guica, T.~Hartman, W.~Song and A.~Strominger, \emph{{The Kerr/CFT
  Correspondence}},
  \href{https://doi.org/10.1103/PhysRevD.80.124008}{\emph{Phys. Rev.}
  {\bfseries D80} (2009) 124008},
  [\href{https://arxiv.org/abs/0809.4266}{{\ttfamily 0809.4266}}].

\bibitem{Compere:2012jk}
G.~Comp\`ere, \emph{{The Kerr/CFT correspondence and its extensions}},
  \href{https://doi.org/10.1007/s41114-017-0003-2}{\emph{Living Rev. Rel.}
  {\bfseries 15} (2012) 11}, [\href{https://arxiv.org/abs/1203.3561}{{\ttfamily
  1203.3561}}].

\bibitem{Hartman:2008pb}
T.~Hartman, K.~Murata, T.~Nishioka and A.~Strominger, \emph{{CFT Duals for
  Extreme Black Holes}},
  \href{https://doi.org/10.1088/1126-6708/2009/04/019}{\emph{JHEP} {\bfseries
  04} (2009) 019}, [\href{https://arxiv.org/abs/0811.4393}{{\ttfamily
  0811.4393}}].

\bibitem{Carlip:1998wz}
S.~Carlip, \emph{{Black hole entropy from conformal field theory in any
  dimension}}, \href{https://doi.org/10.1103/PhysRevLett.82.2828}{\emph{Phys.
  Rev. Lett.} {\bfseries 82} (1999) 2828--2831},
  [\href{https://arxiv.org/abs/hep-th/9812013}{{\ttfamily hep-th/9812013}}].

\end{thebibliography}\endgroup



\providecommand{\href}[2]{#2}\begingroup\raggedright\endgroup

\end{document}